\documentclass[aps,twocolumn,groupedaddress,showpacs]{revtex4}

\usepackage{epsfig,amsbsy,bm}
\newcommand{\eref}[1] {(\ref{#1})}
\newcommand{\Eref}[1] {Eq.~(\ref{#1})}
\newcommand{\Fref}[1] {Fig. \ref{#1}}

\newcommand{\be}{\begin{equation}}
\newcommand{\ee}{\end{equation}}

\renewcommand{\bbox}{\boldsymbol }

\newcommand{\vs}{\vspace*}

\newcommand{\bt}{\begin{tabular}}
\newcommand{\et}{\end{tabular}}
\newcommand{\bp}{\begin{minipage}}
\newcommand{\ep}{\end{minipage}}

\newcommand{\br}{\begin{eqnarray*}}
\newcommand{\er}{\end{eqnarray*}}
\newcommand{\ba}{\begin{eqnarray}}
\newcommand{\ea}{\end{eqnarray}}
\renewcommand{\r}{{\bbox r}}

\renewcommand{\l}{\lambda}

\newcommand{\isum}%
{\mathop{\hbox{$\displaystyle\sum\kern-13.2pt\int\kern1.5pt$}}}

\begin{document}

\bibliographystyle{apsrev}

\title
{Delay in atomic photoionization}

\author{ A. S. Kheifets$^{1,2}$
\footnote[1]{Corresponding author: A.Kheifets(at)anu.edu.au}
}
\author{I. A. Ivanov$^1$}

\affiliation
{$^1$Research School of Physical Sciences,
The Australian National University,
Canberra ACT 0200, Australia
\\
$^2$The Kavli Institute for Theoretical Physics,
University of California
Santa Barbara CA 93106-4030, USA}

\date{\today}

\begin{abstract}
We analyze the time delay between emission of photoelectrons from the
outer valence $ns$ and $np$ sub-shells in noble gas atoms following
absorption of an attosecond XUV pulse. By solving the time dependent
Schr\"odinger equation and carefully examining the time evolution of
the photoelectron wave packet, we establish the apparent ``time zero''
when the photoelectron leaves the atom.  Various processes such as
elastic scattering of the photoelectron on the parent ion and
many-electron correlation affect the quantum phase of the dipole
transition matrix element, the energy dependence of which defines the
emission timing.  This qualitatively explains the time delay between
photoemission from the $2s$ and $2p$ sub-shells of Ne as determined
experimentally by attosecond streaking [{\em Science} {\bf 328}, 1658
(2010)]. However, with our extensive numerical modeling, we were only
able to account for less than a half of the measured time delay of
$21\pm5$~as. We argue that the XUV pulse alone cannot produce such a
large time delay and it is the streaking IR field that is most likely
responsible for this effect.
\end{abstract}


\maketitle

The attosecond streaking has made experimentally accessible the
characteristic timescale of electron motion in atoms
\cite{Baltuska2003,Kienberger2004}.  Among other spectacular
applications of this technique, it has become possible to determine
the time delay between subjecting an atom to a short laser pulse and
subsequent emission of the photoelectron. 
In a recent work by \citet{P.Eckle12052008}, the helium atom was subjected
to a near-infrared laser pulse with intensity of several units of
$10^{14}$~W\slash cm$^2$. Such a strong field ionization regime could
be characterized by a fairly small Keldysh parameter
$\gamma\simeq1$. The time delay in such a photoemission process can be
conveniently analized in terms of nonadiabatic tunneling
\cite{PhysRevA.63.033404}. In a subsequent experiment by
\citet{M.Schultze06252010}, the time delay was measured in neon  in
the XUV photon energy range by high-order harmonic conversion of the
driving near-infrared laser pulse. In this regime, which is
characterized by a moderate intensity, short wavelength and
$\gamma\gg1$, it is believed that the formation of the outgoing wave
packet follows instantaneously temporal variation of the incident
electromagnetic field. Nevertheless, a sizable time delay of
$21\pm5$~as was reported between photoemission from the $2s$ and $2p$
valence sub-shells of Ne. \citet{M.Schultze06252010} argued that a
comprehensive temporal characterization of photoemission on the
attosecond time scale could provide a new insight into intra-atomic
electron correlations. Indeed, the best theoretical treatment within
an independent electron model could only account for 4.0~as time
delay. When the theoretical model was corrected for electron
correlations before and after photoionization, a relative delay of
6.4~as was obtained. This unresolved difference between the measured
and calculated time delays puts many-electron models of atomic
photoionization under significant strain. If substantiated, this
difference could potentially point to new physical mechanisms
underpinning electromagnetic interaction in atoms on the attosecond
time scale.

In this Letter, we perform extensive study of the time delay between
the $ns$ and $np^2$ outer valence sub-shell photoionization in noble
gas atoms. We employ both the explicit time-dependent and stationary
treatment of the photoionization process. To this end, we solve the
time-dependent Schr\"oding equation (TDSE) in the single active
electron  approximation. By carefully examining the time
evolution of the photoelectron wave packet, we establish the apparent
``time zero'' when the photoelectron leaves the atom. To account for
electron correlation, we solve a set of coupled integral equations in
the random phase approximation with exchange (RPAE) \cite{A90}. A
similar approximation is employed to account for a possible target
polarization effect \cite{Amusia1974387}. 

Within an independent electron approximation, the time delay is caused
by the energy dependence of the elastic scattering phase shifts of the
photoelectron moving in the Hartree-Fock (HF) potential of the Ne$^+$
ion. The many-electron correlation, which is due to inter-shell
$2s$-$2p$ coupling, depends on the energy of the photon and thus add
additional component to the quantum phase of the dipole matrix
element.  Both these effects account for the time delay not exceeding
$\sim 10$~as. This recovers only about one half of the experimental
value of $21\pm5$~as. We carefully examine other correlation and
polarization corrections, but find them unable to produce any sizable
contribution to the measured time delay. We also analyze these effects
in other noble gas atoms.

\begin{figure}[h]
\vs{4cm}
\epsfxsize=7cm
\epsffile{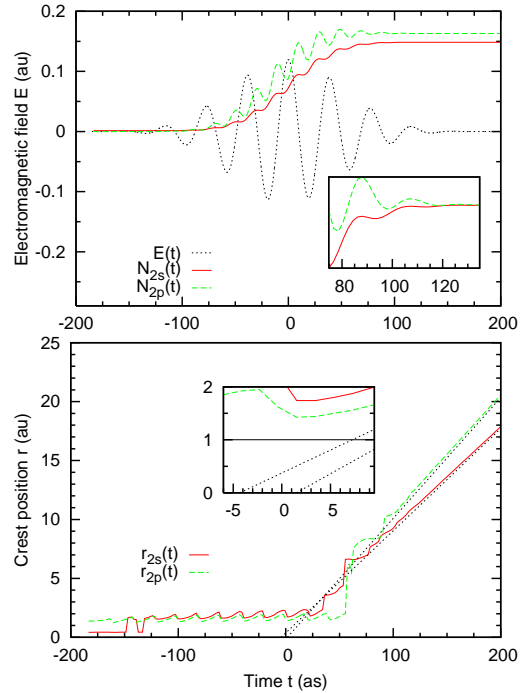}

\caption{
Top: the norm of the wave packets $N(t)$ (scaled arbitrarily) emitted
from the $2s$ and $2p$ sub-shells is plotted as a function of time with
the red solid and green dashed lines, respectively.  The XUV pulse is
over-plotted with the black dotted line. In the inset, the norm variation
$[N(t)-N(T_1)]\slash N(T_1)$ is shown on an expanded time scale near
the pulse end.
Bottom: the crest position of the $2s$ and $2p$ wave packets is shown
with the same linestyles. The crest position after the pulse end is
fitted with the straight line which corresponds to the free
propagation.  In the inset, extrapolation of the free propagation
inside the atom is shown.
\vs{-5mm}
}
\label{Fig1}
\end{figure}

The time-dependent calculation of photoionization in Ne was performed
by radial grid integration of the TDSE using the matrix iteration
method \cite{PhysRevA.60.3125}.  We employed a one-electron basis in
an parametrized optimized effective potential \cite{Sarsa2004163}.
We used the linearly polarized XUV pulse ${\cal E}(t)=E_0 \,g(t) \sin{\omega
t}$ with the envelope $g(t)$ represented by the Nutall window function
and centered at $t=0$. The following field parameters were chosen:
$E_0=0.119$ a.u. (corresponding to the peak intensity of $5\times
10^{14}$ W/cm$^2$), $\omega=106$~eV, $T=2\pi\slash\omega=39$~as and
FWHM=182~as.  Experimental field intensity was not reported by
\citet{M.Schultze06252010}.  Given a typical high-order harmonic
conversion efficiency of $10^{-6}$ \cite{E.Goulielmakis06202008}, the
presently chosen XUV field strength is most certainly larger than the
one used experimentally. With this choice, our calculation is
guaranteed to account for non-perturbative, with respect to the field,
ionization effects if this effects were to be sizable. Other XUV field
parameters used in the present work were identical to those used in
the experiment.  The XUV pulse described above is
shown on the top panel of \Fref{Fig1} (dotted black line). The pulse
is truly off outside the interval $\pm T_1$, where $T_1\simeq 5T$
which is about twice the FWHM.

The solution of the TDSE satisfies the initial condition
$\Psi(\r,t<T_1)=\phi_i(\r)$ which corresponds to a bound electron
state on the atomic shell $i$ to be ionized.  So the shell index $i$ is
implicit in the following but omitted for brevity. 
The wavepacket representing the photoelectron ejected from a given
shell  is defined as
\begin{equation}
\Phi(\r,t)={\scriptstyle \sum_L \int} a_{kL}(t) \chi_{kL}(\r) e^{-i E_k t}\ dk \ ,
\label{1}
\end{equation}
where $ a_{kL}(t)=e^{iE_k t} \langle \chi_{kL}|\Psi(t) \rangle$ are
the projection coefficients of the solution of the TDSE on the
continuum spectrum of the atom. The continuum state
$\chi_{kL}(\r)=R_{kl}(r) Y_L(\r\slash r)$ is the product of the radial
orbital with the asymptotic
$
R_{kl} \propto
\sin{\big[kr+\delta_l(k)+{1\slash k}\ln(2kr)-l\pi\slash 2}
\big]
$
and the spherical harmonic $Y_L(\r\slash r)$ with $L\equiv l,m$. 
%
%
The projection coefficients $a_{kl}(t)$ cease to depend on time for
$|t|>T_1$ when the driving XUV pulse is off.

There are two convenient indicators of the evolution
of the wave packet~\eref{1}. One is the norm given by the integral
$N(t)=\sum_L \int  dk \ |a_{kL}(t)|^2$. This norm is plotted on the
top panel of \Fref{Fig1} with the red solid and green dashed lines for
the wave packets originated from the $2s$ and $2p$ sub-shells,
respectively. For better clarity, these curves are scaled and over-plotted
on the electromagnetic pulse.

The figure shows clearly that the evolution of the $2s$ and $2p$ wave
packets starts and ends at the same time without any noticeable
delay. This is further visualized in the inset where the variation of
the norm $[N(t)-N(T_1)]\slash N(T_1)$ is plotted on an expanded time
scale near the driving pulse end.  Indeed, the norm starts deviating
from zero with the  rise of the XUV pulse and reaches its
asymptotic value once the interaction with the XUV pulse is over.

Another marker of the wavepacket dynamics is the crest
position, defined as a location of the global maximum of the electron
density. The latter quantity is truly informative only when the
electron is outside the atom and the wavepacket is fully formed, having
one well-defined global maximum.
On the bottom panel of \Fref{Fig1}, we show the crest position of the
$2s$ and $2p$ wave packets propagating in time. This figure can be
viewed as a more realistic version of a somewhat idealized and
simplified graph presented in Fig.~1 of \citet{M.Schultze06252010}.
We see that evolution of the norm and the movement of the crest
commence and cease at about the same time.

The movement of the crest becomes almost linear when the norm reaches
its asymptotic value and the wave packet is fully formed. Once fitted
with the linear time dependence $r=k(t-t_0)+r_0$ for large times
$t>T_1$ (shown as a dotted straight line) and back propagated inside
the atom, the $2s$ wavepacket seems to have an earlier start time
$t_0$ than that of the $2p$ wavepacket. This difference is magnified
in the inset. It is about $6$~as at the origin $r_0=0$ and about 4~as
at the distance $r_0\sim 1$~au which corresponds to the size of the
valence shell of the Ne atom. We see that at the origin $t_0^{2s}<0$
and $t_0^{2p}>0$ are shifted to the opposite direction with respect to
the peak of the driving XUV pulse which sets the start time   of the
photoionization process. Thus the seeming (or apparent according to
Ref.~\cite{M.Schultze06252010}) ``time zero'' of the wave packet,
which is inferred by the backward time propagation, is different from
the physical (or real) ``time zero'' $t=0$.

The origin of this shift is most clearly elucidated within the
perturbation theory (PT) framework, which should be applicable under
the present field conditions of a single photon transition with 
$\gamma\gg1$ \cite{Delone1985}. Under these conditions,
%
\begin{equation}
a_{kL}(t>T_1)=-i 
\int_{-\infty}^{\,\infty} \langle
\chi_{kL}|z|\phi_i\rangle e^{i(E_k-\epsilon_i)\tau} 
{\cal E}(\tau)\ d\tau \ .
\label{2}
\end{equation}
Here we extended the integration limits outside the pulse duration and
wrote the dipole matrix element
$
\langle \chi_{k\l}|z|\phi_i\rangle
$ 
 in the length gauge. By separating the angular and radial
integration, we can present this matrix element in the reduced form
\be
\langle \chi_{kL}|z|\phi_i\rangle \propto 
C^{lm}_{10\,l_im_i}
d_\l(k) \ ,
\label{3}
\ee
where $C^{lm}_{10\,l_im_i}$ is the Clebsch-Gordan coefficient,
$\l\equiv l,i$ and $d_\l(k)$ is real.
With this definition, we can write $a_{kL}\propto-i\,d_\l(k)
\tilde {\cal E}(E_k-\epsilon_i)$, where the Fourier transform of the XUV field
$
\tilde {\cal E}(\omega)=
\int_{-\infty}^{\infty} e^{i\omega\tau} {\cal E}(\tau)\ d\tau
$
is  real for a symmetric pulse that we presently consider.
We note that Eqs.~\eref{1}--\eref{3} are equivalent to Eqs.~(S5)--(S8)
of \citet{M.Schultze06252010} given in their supporting online
material.

To describe the motion of the wave packet \eref{1}, we apply the usual
saddle-point method. For each $l$, the crest of the wave packet is
moving at large times $t>T_1$ quasi-classically along the trajectory
which is given by the  equation:
\begin{equation}
r
=
k\Big\{
t-{d\over dE} \big[\delta_l(k)+{1\over k} \ln(2kr) 
\big]_{k=\sqrt{2E_0}}
\Big\} \ .
\label{4}
\end{equation}
Since the logarithm is a slowly varying function which can be absorbed
into a constant, \Eref{4} describes a straight line $r=k(t-t_0)+r_0$
with $t_0=d\delta_l(k)\slash dE\big|_{k=\sqrt{2E_0}}\ $. Thus the
relative time delay between various photoionization channels is
determined primarily by the derivatives of the corresponding
elastic scattering phases \cite{deCarvalho200283}.

\begin{figure}[h]
\epsfxsize=6.5cm
\epsffile{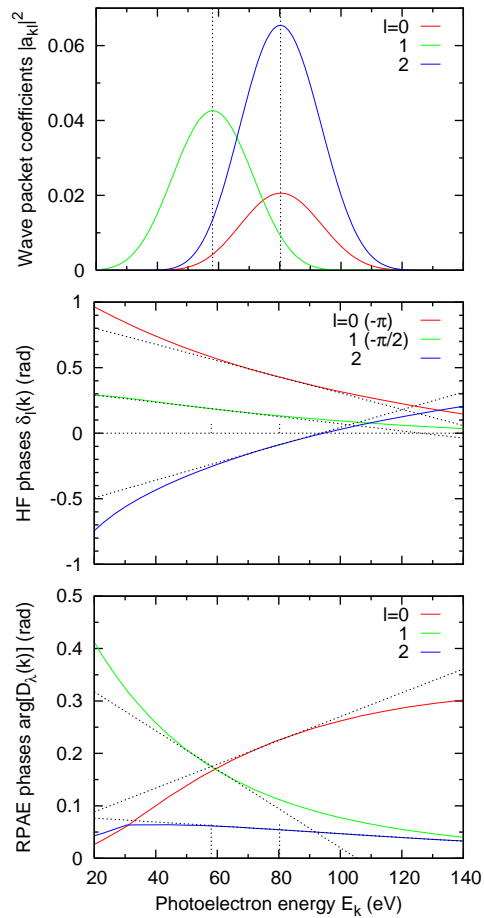}

\vs{7.5cm}

\caption{
Top: Expansion coefficients $|a_{kl}|^2$ plotted versus the
photoelectron energy $E_k=k^2\slash2$ which is expressed in eV. Middle:
the HF scattering phases. Bottom: phases of the RPAE dipole matrix
elements $\arg[D_\lambda(k)]$.  
\vs{-5mm}}
\label{Fig2}
\end{figure}

The scattering phases $\delta_l(k)$ of the photoelectron moving in the
field of a singly charged Ne$^+$ ion are shown on the middle panel of
\Fref{Fig2}.  The photoelectron ejected from the $2s$ shell has
only one value of the angular momentum $l=0$ whereas the $2p$
photoelectron can acquire two angular momenta $l=0$ and 2.  The phases
in the $s$- and $p$-waves are shifted downwards by $\pi$ and
$\pi\slash2$, respectively, for better clarity.  On the top panel of
the same figure, we display the asymptotic projection coefficients
$a_L(k)$ for $l=0$, 1 and 2 and $m=0$. 
The centers of the energy distribution of the $l$-projected
coefficients (indicated by the vertical dotted lines on the top panel)
define the position of the energy derivative of the corresponding
scattering phases
$
d\delta_l(k)\slash dE\Big|_{k=\sqrt{2E_0}}
$
(indicated by the straight lines on the middle panel).  We see that
the energy derivatives of the $s$- and $p$-phases are negative whereas
that of the $d$-phase is positive. This is so due to the presence of
the occupied $s$ and $p$ states in the Ne$^+$ ion which disturbs
otherwise monotonic increase with energy of the Coulomb phase (the
Levinson-Seaton theorem \cite{PhysRevA.52.3824}).  Because the $2p\to
kd$ transition is strongly dominant over the $2p\to ks$ one, as is
seen from the corresponding projection coefficients on the top panel
of \Fref{Fig2}, it is the $d$-phase that determines the shift of the
apparent ``time zero'' of the $2p$ wave packet relative the physical
``time zero'' $t=0$. This shift is positive fot the $2p$ wave packet
and negative for the $2s$ one, in accordance with our observation
displayed in the inset of the bottom panel of \Fref{Fig1}.

So far, we confined ourselves with an independent electron
approximation and calculated the dipole matrix elements $d_\l(k)$ and
the scattering phases $\delta_l(k)$ in the HF approximation
\cite{CCR76,CCR79}. It is well known, however, that many-electron
correlation modifies strongly the dipole matrix elements. The full
account for this effect can be taken within the RPAE model \cite{A90}
by solving a set of coupled integral equations
\be
D_\l(k) = d_\l(k)+ \sum_\nu\int\! dp\, D_\nu(p)\ \chi_\nu(p) \
U_{\nu\l}(p,k)
\label{5}
\ee
Here $\chi_\nu(p)=(\omega-E_p-\epsilon_\nu+i\epsilon)^{-1}$ is the
Green's function and $U_{\nu\l}(p,k)$ is the Coulomb interaction
matrix. The one-electron HF basis corresponding to the field of the
singly charged Ne$^+$ ion accounts for direct photoelectron
interaction with its parent shell.  It is therefore the
inter-shell Coulomb interaction with $\nu\ne\l$ that should only be
included into
\Eref{5}.
Since the Green's function is complex, the dipole matrix
elements $D_\l(k)$ acquire an additional phase which is plotted on
the bottom panel of \Fref{Fig2}. 

The HF phase derivatives alone account for the apparent ``time zero''
shift between the $2s$ and $2p$ ionization $\Delta t^{2s-2p}_0 =
6.2$~as. The RPAE correction adds an extra 2.2~as. In total, this
accounts for the apparent ``time zero'' shift $\Delta t^{2s-2p}_0 =
8.4$~as. Both the HF and RPAE phases are smooth functions of the
photoelectron energy and their averaging over the bandwidth of the XUV
pulse does not change these numbers in a noticeable way.  The
analogous values reported by
\citet{M.Schultze06252010} for the independent electron model and the
correlation correction are 4.0~as and 2.4~as, respectively. Both sets
of calculations are quite close and well below the experimental value
of $21\pm5$~as.

One could argue that the complete account for many-electron
correlation within the TDSE, rather than adding this correlation {\em
ad hoc}, could modify the present result. This is, however, unlikely
given the nature of the RPAE which is a direct generalization of the
HF method in the presence of an oscillatory external electromagnetic
field \cite{Thouless1972}. The only approximation taken when deriving
\Eref{5} is that at any instant of time the atomic wave function is an
anti-symmetric product of one-electron functions.  It is quite a
robust approximation under the field parameters considered above.

We also evaluated the time delay of the wave packet relative to the
XUV pulse in other noble gases. In He, the wave packet emitted from
the $1s$ shell is delayed by $\sim2$~as relative to the center of the
XUV pulse. This follows from the independent electron HF calculation
which returns a positive derivative of the $p$ phase shift as there is
no occupied $p$ orbital in the He$^+$ ion. It is also confirmed by the
correlated convergent close-coupling (CCC) model which is known to
produce benchmark photoionization results for He in the XUV range
\cite{0953-4075-35-15-201}. It is to be compared with 5~as delay
reported for He by \citet{M.Schultze06252010}. In heavier noble gases,
Ar and Kr, the difference of the HF $p-$ and $d-$phase derivatives
becomes smaller as occupation of the ionic orbitals increases in line
with the Levinson-Seaton theorem. In Kr, the $d-$phase derivative
becomes negative as the $3d$ orbital is occupied. Accordingly, the
time delay between the wave packets emitted from the $ns$ and $np$
valence sub-shells is getting smaller.  When the HF and RPAE phase
derivatives are combined, it results in 5.8~as delay in Ar and nearly
zero delay in Kr around the 100~eV photon energy mark.

In conclusion, we examined various effects leading to the shift
between the apparent ``time zero'' of the photoelectron wave packets
emitted from the $2s$ and $2p$ shells in neon relative to the center
of the XUV pulse which sets the timing of the photoionization process.
We found that this shift is primarily due to the energy derivative of
the HF elastic scattering phase shifts which differs significantly for
various partial waves. The RPAE correction, which accounts for
many-electron correlation, is rather small and cannot explain the
profound difference between the theoretical and experimental time
delay.

The apparent ``time zero'' is only meaningful when the wave packet is
detected at large distances from the atom as in attosecond streaking
experiments. This apparent ``time zero'' has little to do with the
real time when the atomic photoionization begins which is fully
determined by the driving XUV pulse alone. In this sense, the
attosecond streaking is not informative on the early stages of the
photoionization process. However, this technique allows one to determine
the energy derivative of the quantum phase of the dipole matrix
element \cite{PhysRevLett.105.073001}, thus facilitating the so-called
complete photoionization experiment \cite{0953-4075-37-6-010}. This is
particularly important in those targets where the many-electron
correlation is significant.

The full potential of the attosecond streaking technique and
its successful application in atomic collision physics can only be
realized if the current strong disagreement between theory and
experiment in Ne is resolved. The present study was not able to do so.
Our simulations and analytic arguments indicate that the XUV pulse
alone cannot produce such a large time delay and it is the streaking
IR field that is most likely responsible for this effect.

The authors acknowledge support of the Australian Research Council in
the form of the Discovery grant DP0771312.  Resources of the National
Computational Infrastructure (NCI) Facility were employed. One of the
authors (ASK) wishes to thank the Kavli Institute for Theoretical
Physics for hospitality.  This work was supported in part by the NSF
Grant No.~PHY05-51164
\vs{-5mm}


\end{document}